\documentclass[11pt,a4paper]{article}
\usepackage{amsmath}
\usepackage[mathcal]{eucal}
\usepackage{latexsym}
\usepackage{amssymb}

\begin{titlepage} 
\title{General covariance for a proposal for 4-D gravity}
\author{Eyo Eyo Ita III}
\medskip
\input amssym.def
\input amssym.tex

\def \in{\indent}

\begin{document}
\maketitle
\bigskip
\centerline{Department of Applied Mathematics and Theoretical Physics} 
\smallskip
\centerline{Centre for Mathematical Sciences, University of Cambridge, Wilberforce Road}
\smallskip
\centerline{Cambridge CB3 0WA, United Kingdom}
\smallskip
\centerline{eei20@cam.ac.uk} 
    
\bigskip
  
\begin{abstract}
In this paper we extend some previous results for a new proposal for gravity and place them into overall context.  The basic fields of this proposal provide an off-shell realization of symmetry with respect to SO(3,C) gauge transformations and general coordinate transformations of 4-dimensional spacetime.
\end{abstract}
\end{titlepage}

\section{Introduction}

In this paper we review and extend some results concerning a proposal for a new description of gravity named the instanton representation of Plebanski gravity.  We have approached this description from the standpoint of general covariance as an alternative to the standard canonical approach to general relativity.  We place the results obtained thus far into their overall context, with a view to addressal of the quantum theory in future works.  The main idea is that a theory of four dimensional gravity should exhibit invariance under gauge transformations and general coordinate transformations, an invariance which should manifestly be preserved under the canonical formalism.  The organization of this paper is as follows.  First we recount the main results of \cite{EYOITA} and \cite{EYOITA1}, setting the stage for the present paper.  In section 2 we extend the relevant symmetry group from gauge transformations and spatial diffeomoprhisms to include the full spacetime general coordinate transformations.  In conjunction we demonstrate the consistency of this by off-shell closure of the algebra on all of the basic fields of the theory.  In section 3 we recount the relation to general relativity, and in section 4 we provide a conclusion.\par
\subsection{Setting the stage}
Let $M$ be a four-dimensional spacetime manifold.  The set of general coordinate transformations 
\begin{eqnarray}
\label{GENCOORD}
x^{\mu}\rightarrow{x^{\prime}}^{\mu}={x}^{\mu}+\xi^{\mu}(x),
\end{eqnarray}
\noindent
referred to as $Diff(M)$, induces the following Lie algebra between any two smooth vector fields $\xi,\zeta\in{C}^{\infty}(M)$, given by
\begin{eqnarray}
\label{LIE}
\bigl[\xi^{\mu}\partial_{\mu},\zeta^{\nu}\partial_{\nu}\bigr]=\bigl(\xi^{\mu}\partial_{\mu}\zeta^{\nu}-\zeta^{\mu}\partial_{\mu}\xi^{\nu}\bigr)\partial_{\nu}.
\end{eqnarray}
\noindent
We would like to propose a theory of gravity invariant under (\ref{GENCOORD}).  The basic fields of the associated action $I_{Inst}$ should provide a realization of (\ref{LIE}) independently of any equations of motion or canonical structure, and should thus constitute an off-shell realization.  For the basic fields we will use a $SO(3,C)$ gauge connection $A^a_{\mu}$ and a 3 by 3 matrix $\Psi_{ae}$ taking its values in two copies of $SO(3,C)$.\footnote{For index conventions, the Latin symbols $a,b,c,\dots$ will assume values $1-3$ and will refer to internal $SO(3,C)$ indices, and the Greek symbols $\mu,\nu,\dots$ taking values $0-3$ will refer to spacetime indices.}  The proposed action is given by
\begin{eqnarray}
\label{ACTIONINSTTT}
I_{Inst}=\int{dt}\int_{\Sigma}d^3x\Bigl(\Psi_{ae}B^i_e\dot{A}^a_i+A^a_0B^i_eD_i\Psi_{ae}\nonumber\\
+\epsilon_{ijk}N^iB^j_aB^k_e\Psi_{ae}-iN(\hbox{det}B)^{1/2}\sqrt{\hbox{det}\Psi}\bigl(\Lambda+\hbox{tr}\Psi^{-1}\bigr)\Bigr),
\end{eqnarray}
\noindent
where $\Sigma$ represents 3-dimensional spatial hypersurfaces which foliate spacetime into $M=\Sigma\times{R}$.  We have defined $B^i_a={1 \over 2}\epsilon^{ijk}F^a_{jk}$ as the magnetic field of $A^a_i$, which is the spatial 
part of $A^a_{\mu}$.  This constitutes the spatial part of the curvature of $A^a_{\mu}$, given by
\begin{eqnarray}
\label{ASSUME1}
F^a_{\mu\nu}=\partial_{\mu}A^a_{\nu}-\partial_{\nu}A^a_{\mu}+f^{abc}A^b_{\mu}A^c_{\nu}
\end{eqnarray}
\noindent
where $f^{abc}=\epsilon^{abc}$ are the $SO(3,C)$ structure constants.\par
\indent
In addition to being invariant under (\ref{GENCOORD}), the proposed theory should be invariant also under $SO(3,C)$ gauge transformations.  Define $SO(3,C)*Diff$ as the set of all spacetime general coordinate 
transformations and $SO(3,C)$ gauge transformations continuously connected to the identity.  Under an infinitesimal $SO(3,C)$ gauge transformation $\delta_{\vec{\eta}}$, the connection $A^a_{\mu}$ transforms as \cite{GRAV}
\begin{eqnarray}
\label{COMPLETENESS}
\delta_{\vec{\eta}}A^a_{\mu}=-D_{\mu}\eta^a=-\partial_{\mu}\eta^a-f^{abc}A^b_{\mu}\eta^c,
\end{eqnarray}
\noindent
where $f^{abc}=\epsilon^{abc}$ are the $SO(3,C)$ structure constants.  Under infinitesimal spacetime diffeomorphisms $\delta_{\xi}$, the connection $A^a_{\mu}$ transforms according to the Lie derivative 
\begin{eqnarray}
\label{COMPLETENESS5}
\delta_{\xi}A^a_{\mu}=\xi^{\nu}\partial_{\nu}A^a_{\mu}+(\partial_{\mu}\xi^{\nu})A^a_{\nu}.
\end{eqnarray}
\noindent
It has been shown in \cite{EYOITA} that $SO(3,C)*Diff$ forms a Lie algebra, which closes on the field $A^a_{\mu}$
\begin{eqnarray}
\label{FINALRESULTS}
\bigl[\delta_{\vec{\theta}},\delta_{\vec{\eta}}\bigr]A^a_{\mu}=-\delta_{\vec{\theta}\times\vec{\eta}}A^a_{\mu};~~\bigl[\delta_{\xi},\delta_{\vec{\eta}}\bigr]A^a_{\mu}=-\delta_{(\delta_{\xi},\vec{\eta})}A^a_{\mu};~~
\bigl[\delta_{\xi},\delta_{\zeta}\bigr]A^a_{\mu}=-\delta_{[\xi,\zeta]}A^a_{\mu}.
\end{eqnarray}
\noindent
For the field $\Psi_{ae}$ only the spatial part of the algebra, $SO(3,C)*diff\subset{SO}(3,C)*Diff$, has been shown to close in \cite{EYOITA}.\par
\indent
In this paper we will extend the algebra on $\Psi_{ae}$ to include the temporal parts of (\ref{GENCOORD}).  Recall from \cite{EYOITA} that $\Psi_{ae}$ has been shown to form an off-shell realization of the $SO(3,C)$ part of the algebra.  Hence to extend this to $SO(3,C)*Diff$ it suffices to show that the algebra (\ref{LIE}) closes on $\Psi_{ae}$, and in addition forms a Lie algebra with $SO(3,C)$ which also closes on $\Psi_{ae}$.  The transformation 
of $\Psi_{ae}$ under an infinitesimal $SO(3,C)$ transformations parametrized by a $SO(3,C)$-valued 3-vector is given by \cite{EYOITA}
\begin{eqnarray}
\label{RESUULTI}
\delta_{\vec{\eta}}\Psi_{ae}=f_{aec}^{bf}\Psi_{bf}\eta^c,
\end{eqnarray}
\noindent
where we have defined
\begin{eqnarray}
\label{DEFINED}
f_{aec}^{bf}=f_{abc}\delta_{ef}+f_{efc}\delta_{ab}.
\end{eqnarray}
\noindent
As mentioned, in \cite{EYOITA} we have established $\delta_{\vec{N}}\Psi_{ae}=N^i\partial_i\Psi_{ae}+(\partial_iN^i)\Psi_{ae}$ for any spatial 3-vector $N^i$.\footnote{This is an error which we will correct in the present paper.  The transformation should be given by $\delta_{\vec{N}}\Psi_{ae}=N^i\partial_i\Psi_{ae}$, which signifies that $\Psi_{ae}$ transforms as a scalar as opposed to a scalar density of weight one.  This correction does not affect the final results or conclusions in \cite{EYOITA}.}  In the present paper we will extend this to include the temporal transformations, thus extending this to
\begin{eqnarray}
\label{DEFINED1}
\delta_{\xi}\Psi_{ae}=\xi^{\sigma}\partial_{\sigma}\Psi_{ae}
\end{eqnarray}
\noindent
for any 4-vector $\xi^{\sigma}$.

\section{Extension of transformation properties}

In order to proceed, we must show two things, namely (i) that transformations (\ref{DEFINED1}) form a closed algebra on $\Psi_{ae}$.  This is guaranteed, since (\ref{LIE}) is an identity when acting on any coordinate scalar.  The 
field $\Psi_{ae}$ is a coordinate scalar since it does not have any spacetime indices.  Hence it suffices to act on $\Psi_{ae}$ with both sides of (\ref{LIE}) in order to see that this is the case. (ii) We must show 
that (\ref{GENCOORD}) and the transformations (\ref{RESUULTI}) close on $\Psi_{ae}$, which entails finding the commutator of the two transformations
\begin{eqnarray}
\label{WHICHENTAILS}
\bigl[\delta_{\xi},\delta_{\vec{\eta}}\bigr]\Psi_{ae}=\delta_{\xi}(\delta_{\vec{\eta}}\Psi_{ae})-\delta_{\vec{\eta}}(\delta_{\xi}\Psi_{ae})
=\delta_{\xi}(f_{aec}^{bf}\Psi_{bf}\eta^c)-\delta_{\vec{\eta}}(\xi^{\sigma}\partial_{\sigma}\Psi_{ae})
\end{eqnarray}
\noindent
where we have used (\ref{RESUULTI}) and (\ref{DEFINED1}).  Proceeding from (\ref{WHICHENTAILS}), and using the fact that the variations act on the fields, we have
\begin{eqnarray}
\label{WHICHENTAILS1}
f_{aec}^{bf}(\delta_{\xi}\Psi_{bf})\eta^c-\xi^{\sigma}\partial_{\sigma}(\delta_{\vec{\eta}}\Psi_{ae})\nonumber\\
=f_{aec}^{bf}(\xi^{\sigma}\partial_{\sigma}\Psi_{bf})\eta^c-\xi^{\sigma}\partial_{\sigma}(f_{aec}^{bf}\Psi_{bf}\eta^c)\nonumber\\
=-f_{aec}^{bf}\Psi_{bf}(\xi^{\sigma}\partial_{\sigma}\eta^c)=-f_{aec}^{bf}\Psi_{bf}(L_{\xi}\eta^c),
\end{eqnarray}
\noindent
where $L_{\xi}\eta^c=\xi^{\sigma}\partial_{\sigma}\eta^c\equiv\delta_{\xi}\eta^c$ is the Lie derivative of $\eta^c$ along the vector field generating the flow $\xi^{\sigma}$.  The result is that
\begin{eqnarray}
\label{WHICHENTAILS2}
\bigl[\delta_{\xi},\delta_{\vec{\eta}}\bigr]\Psi_{ae}=-\delta_{\delta_{\xi}\vec{\eta}}\Psi_{ae}.
\end{eqnarray}
\noindent
The result is that the algebra (\ref{FINALRESULTS}) extends to the field $\Psi_{ae}$.  Let us rewrite the algebra for completeness
\begin{eqnarray}
\label{WAVEITER}
\bigl[\delta_{\vec{\theta}},\delta_{\vec{\eta}}\bigr]A^a_{\mu}=-\delta_{\vec{\theta}\times\vec{\eta}}A^a_{\mu};~~
\bigl[\delta_{\vec{\theta}},\delta_{\vec{\eta}}\bigr]\Psi_{ae}=-\delta_{\vec{\theta}\times\vec{\eta}}\Psi_{ae}\nonumber\\
\bigl[\delta_{\xi},\delta_{\vec{\eta}}\bigr]A^a_{\mu}=-\delta_{\delta_{L_{\xi}}\vec{\eta}}A^a_{\mu};~~
\bigl[\delta_{\xi},\delta_{\vec{\eta}}\bigr]\Psi_{ae}=-\delta_{L_{\xi}\vec{\eta}}\Psi_{ae}\nonumber\\
\bigl[\delta_{\xi},\delta_{\zeta}\bigr]A^a_{\mu}=-\delta_{[\xi,\zeta]}A^a_{\mu};~~
\bigl[\delta_{\xi},\delta_{\zeta}\bigr]\Psi_{ae}=-\delta_{[\xi,\zeta]}\Psi_{ae}.
\end{eqnarray}
\noindent
The significance of this result is that the purely temporal part of the algebra of (\ref{LIE}) forms a subalgebra with respect to the field $\Psi_{ae}$ in addition to $A^a_{\mu}$.\par
\indent
Another interesting relation arising from this result is that one can, using $\Psi_{ae}$ and $A^a_{\mu}$, construct a quantity $\Sigma^a_{\mu\nu}$ antisymmetric in $\mu$ and $\nu$ given by
\begin{eqnarray}
\label{ASSUME}
\Sigma^a_{\mu\nu}=\Psi_{ae}F^e_{\mu\nu}.
\end{eqnarray}
\noindent
The transformation properties of $A^a_{\mu}$ under (\ref{GENCOORD}) imply the the following transformation properties for the curvature $F^a_{\mu\nu}$ as derived in Appendix A
\begin{eqnarray}
\label{ASSUME2}
\delta_{\xi}F^a_{\mu\nu}=\xi^{\sigma}\partial_{\sigma}F^a_{\mu\nu}+(\partial_{\mu}\xi^{\sigma})F^a_{\sigma\nu}+(\partial_{\nu}\xi^{\sigma})F^a_{\mu\sigma}.
\end{eqnarray}
\noindent
Application of the Liebniz rule to (\ref{ASSUME}) yields
\begin{eqnarray}
\label{ASSUME5}
\delta_{\xi}\Sigma^a_{\mu\nu}=(\delta_{\xi}\Psi_{ae})F^e_{\mu\nu}+\Psi_{ae}\delta_{\xi}F^e_{\mu\nu}.
\end{eqnarray}
\noindent
Substitution of (\ref{ASSUME2}) and (\ref{DEFINED1}) into (\ref{ASSUME5}) yields the following transformation property of $\Sigma^a_{\mu\nu}$
\begin{eqnarray}
\label{ASSUME3}
\delta_{\xi}\Sigma^a_{\mu\nu}=\xi^{\sigma}\partial_{\sigma}\Sigma^a_{\mu\nu}+(\partial_{\mu}\xi^{\sigma})\Sigma^a_{\sigma\nu}+(\partial_{\nu}\xi^{\sigma})\Sigma^a_{\mu\sigma},
\end{eqnarray}
\noindent
which is consistent with what one expects of a second-rank tensor.  The spatial restriction of (\ref{ASSUME}) is given by
\begin{eqnarray}
\label{RESTTR}
\widetilde{\sigma}^i_a=\Psi_{ae}B^i_e
\end{eqnarray}
\noindent
where $\widetilde{\sigma}^i_a={1 \over 2}\epsilon^{ijk}\Sigma^a_{jk}$ which plays the role of a densitized triad in the Ashtekar formulation of general relativity \cite{ASH1}.

\section{Relation to general relativity}

The Hamiltonian constraint in the canonical treatment of general relativity is the generator of temporal evolution.  Our proposition is that for (\ref{ACTIONINSTTT}), this should provide a canonical realization of the 
temporal part of (\ref{LIE}).  The Hamiltonian constraint can be read off directly from (\ref{ACTIONINSTTT}) as
\begin{eqnarray}
\label{HAMILL}
H[N]=\int_{\Sigma}d^3xN(\hbox{det}B)^{1/2}\sqrt{\Psi}\bigl(\Lambda+\hbox{tr}\Psi^{-1}\bigr).
\end{eqnarray}
\noindent
Evidence for the validity of this proposition is provided in \cite{EYOITA1} using a reduced version of (\ref{ACTIONINSTTT}), where it is shown that the Hamiltonian constraint forms a closed algebra
\begin{eqnarray}
\label{HAMIL1}
\bigl[H[N],H[M]\bigr]=H\bigl[q^i(M\partial_iN-N\partial_iM)\bigr]
\end{eqnarray}
\noindent
for phase space structure functions $q^i=q^i(\Psi_{ae},A^a_i)$.  This is in contrast to the Teitelboim algebra of Hamiltonian constraints \cite{TEITEL}
\begin{eqnarray}
\label{BOIM}
\bigl[H[N],H[M]\bigr]=H_i\bigl[q^{ij}(M\partial_jN-N\partial_jM)\bigr]
\end{eqnarray}
\noindent
for structure functions $q^{ij}$, where $H_i$ is the diffeomorphism constraint.  The difference is that (\ref{BOIM}) does not close on the Hamiltonian constraint, which implies that a theory of gravity based just on the Hamiltonian constraint cannot be Dirac-consistent except in minisuperspace.  However, the action producing (\ref{HAMIL1}) can be obtained from
\begin{eqnarray}
\label{REPEAT}
I=\int{dt}\int_{\Sigma}d^3x\Bigl(\lambda_1a_2a_3\dot{a}_1+\lambda_2a_3a_1\dot{a}_2+\lambda_3a_1a_2\dot{a}_3\nonumber\\
-iN(\hbox{det}b)^{1/2}\sqrt{\lambda_1\lambda_2\lambda_3}\Bigl(\Lambda+{1 \over {\lambda_1}}+{1 \over {\lambda_2}}+{1 \over {\lambda_3}}\Bigr)
\end{eqnarray}
\noindent
via a simple transformation.  Note that (\ref{REPEAT}) can be seen as (\ref{ACTIONINSTTT}) restricted to $A^a_i=diag(a_1,a_2,a_3)$ and $\Psi_{ae}=diag(a_1,a_2,a_3)$, with the Gauss' law and diffeomorphism constraints removed by hand.  As shown in \cite{EYOITA1}), this action is based only on the Hamiltonian constraint and is Dirac-consistent while having two degrees of freedom per point.  Additionally, the action (\ref{REPEAT}) is not a minisuperspace action since it has spatial derivatives in $(\hbox{det}b)$, where $b^i_a$ is the magnetic field of the diagonal connection $A^a_i$.
\section{Conclusion}
In this paper we have shown that certain fields $\Psi_{ae},A^a_{\mu}$ provide an off-shell realization of the Lie algebra of general coordinate and $SO(3,C)$ gauge transformations $SO(3,C)*Diff$.  We have provided a proposal for an action for general relativity $I_{Inst}$ based on these fields.  We have shown in \cite{EYOITA1} that a reduced form of the action directly obtainable from (\ref{REPEAT}) preserves the subalgebra of temporal transformations, in the sense that two Hamiltonian constraints Poisson-commute into a Hamiltonian constraint.  The action (\ref{REPEAT}) can be obtained from (\ref{ACTIONINSTTT}) by hand,\footnote{A main direction of future research should be to determine whether (\ref{REPEAT}) is some sort of reduced phase space version of (\ref{ACTIONINSTTT}).  At the present stage, this has not yet been conclusively demonstrated.} which on first sight brings into question its relevance to general relativity.  However, it is related to general relativity in at least two respects: (i) The theory (\ref{REPEAT}) uses the same Hamiltonian constraint appearing in GR. (ii) It is a Dirac consistent theory as shown 
in \cite{EYOITA1}, and has two degrees of freedom per point on its reduced phase space.\footnote{So it appears to be some form of reduced gravity where the $SO(3,C)*diff$ portion of the original $SO(3,C)*Diff$ invariant theory has been removed.}\par

\section{Appendix A}

Given (\ref{ASSUME1}) and (\ref{COMPLETENESS5}), we will prove (\ref{ASSUME2}).  First we have the relation
\begin{eqnarray}
\label{PROVE}
\delta_{\xi}F^a_{\mu\nu}=\partial_{\mu}(\delta_{\xi}A^a_{\nu})-\partial_{\nu}(\delta_{\xi}A^a_{\mu})+f^{abc}(\delta_{\xi}A^b_{\mu})+f^{abc}A^b_{\mu}(\delta_{\xi}A^c_{\nu}).
\end{eqnarray}
\noindent
Substituting (\ref{COMPLETENESS5}) into (\ref{PROVE}), we have
\begin{eqnarray}
\label{PROVE1}
\delta_{\xi}F^a_{\mu\nu}=\partial_{\mu}\bigl(\xi^{\sigma}\partial_{\sigma}A^a_{\nu}+A^a_{\sigma}(\partial_{\nu}\xi^{\sigma})\bigr)
=\partial_{\nu}\bigl(\xi^{\sigma}\partial_{\sigma}A^a_{\mu}+A^a_{\sigma}(\partial_{\mu}\xi^{\sigma})\bigr)\nonumber\\
+f^{abc}\bigl(\xi^{\sigma}\partial_{\sigma}A^b_{\mu}+A^b_{\sigma}(\partial_{\mu}\xi^{\sigma})\bigr)A^c_{\nu}
+f^{abc}A^b_{\mu}\bigl(\xi^{\sigma}\partial_{\sigma}A^c_{\nu}+A^c_{\sigma}(\partial_{\nu}\xi^{\sigma})\bigr).
\end{eqnarray}
\noindent
Expanding the partial derivatives in (\ref{PROVE1}) we have
\begin{eqnarray}
\label{PROVE2}
\xi^{\sigma}\partial_{\mu}\partial_{\sigma}A^a_{\nu}+(\partial_{\mu}\xi^{\sigma})(\partial_{\sigma}A^a_{\nu})+(\partial_{\mu}A^a_{\sigma})(\partial_{\nu}\xi^{\sigma})+A^a_{\sigma}(\partial_{\mu}\partial_{\nu}\xi^{\sigma})\nonumber\\
-\xi^{\sigma}\partial_{\nu}\partial_{\sigma}A^a_{\mu}-(\partial_{\nu}\xi^{\sigma})(\partial_{\sigma}A^a_{\mu})-(\partial_{\nu}A^a_{\sigma})(\partial_{\mu}\xi^{\sigma})-A^a_{\sigma}\partial_{\nu}\partial_{\mu}\xi^{\sigma}\nonumber\\
+f^{abc}\xi^{\sigma}(\partial_{\sigma}A^b_{\mu})A^c_{\nu}+f^{abc}A^b_{\sigma}(\partial_{\mu}\xi^{\sigma})A^c_{\nu}
+f^{abc}A^b_{\mu}\xi^{\sigma}(\partial_{\sigma}A^c_{\nu})+f^{abc}A^b_{\mu}A^c_{\sigma}(\partial_{\nu}\xi^{\sigma}).
\end{eqnarray}
\noindent
The terms involving $f^{abc}$ can be combined using the Liebniz rule, and rearranging terms, (\ref{PROVE2}) simplifies to
\begin{eqnarray}
\label{PROVE3}
\delta_{\xi}F^a_{\mu\nu}=\xi^{\sigma}\partial_{\sigma}\bigl(\partial_{\mu}A^a_{\nu}-\partial_{\nu}A^a_{\mu}+f^{abc}A^b_{\mu}A^c_{\nu}\bigr)\nonumber\\
+(\partial_{\mu}\xi^{\sigma})\bigl(\partial_{\sigma}A^a_{\nu}-\partial_{\nu}A^a_{\sigma}+f^{abc}A^b_{\sigma}A^c_{\nu}\bigr)
+(\partial_{\nu}\xi^{\sigma})\bigl(\partial_{\mu}A^a_{\sigma}-\partial_{\sigma}A^a_{\mu}+f^{abc}A^b_{\mu}A^c_{\sigma}\bigr)\nonumber\\
=\xi^{\sigma}\partial_{\sigma}F^a_{\mu\nu}+(\partial_{\mu}\xi^{\sigma})F^a_{\sigma\nu}+(\partial_{\nu}\xi^{\sigma})F^a_{\mu\sigma}.
\end{eqnarray}

\end{document}